\documentclass{article}
\usepackage[a4paper, top=2cm,bottom=2.cm,left=2cm,right=2cm]{geometry}
\usepackage[english]{babel}
\usepackage[T1]{fontenc}
\usepackage{lmodern}
\usepackage[utf8]{inputenc}
\usepackage[scaled=0.95]{helvet}
\usepackage{mathptmx}
\usepackage{amsmath}
\usepackage{amssymb}
\usepackage{xfrac}
\usepackage{algorithm}
\usepackage{algpseudocode}
\usepackage[dvipsnames]{xcolor}
\usepackage[colorlinks, breaklinks, driverfallback=dvipdfm]{hyperref}
\hypersetup{linkcolor = TealBlue, citecolor = black, urlcolor = TealBlue} 
\usepackage[font={small},labelfont={sf,bf}, margin=0cm, indention = 0cm]{caption}
\usepackage{subcaption}
\usepackage{titling}
\usepackage{authblk}
\usepackage{fancyhdr}
\usepackage{setspace}
\usepackage{sectsty}
\usepackage{microtype}
\usepackage{tikz}
\usepackage{pgf}
\usetikzlibrary{math}
\usepackage{balance}
\usepackage[nolists,nomarkers]{endfloat}
\usepackage{multicol}
\usepackage{csquotes}
\usepackage[style=ieee,isbn=false,url=false]{biblatex}
\DeclareMathAlphabet{\mathcal}{OMS}{cmsy}{m}{n}

\DeclareMathOperator*{\argmin}{arg\,min}
\DeclareMathOperator{\diag}{diag}

\pretitle{\centering\sffamily\LARGE\bfseries}
\posttitle{\par}

\predate{}
\postdate{}
\date{}

\allsectionsfont{\sffamily\bfseries\normalsize}

\title{Resolution enhancement, noise suppression, and joint T2* decay estimation in dual-echo sodium-23 MR imaging using anatomically-guided reconstruction}

\author[1,2]{Georg Schramm}
\author[2]{Marina Filipovic}
\author[3]{Yongxian Qian}
\author[3]{Alaleh Alivar}
\author[3]{Yvonne W. Lui}
\author[2]{Johan Nuyts}
\author[1]{Fernando Boada}

\affil[1]{Radiological Sciences Laboratory, Stanford University, School of Medicine, California, US}
\affil[2]{Department of Imaging and Pathology, KU Leuven, Belgium}
\affil[3]{Center for Biomedical Imaging, Department of Radiology, New York University (NYU) Grossman School of Medicine, New York, US}

\begin{document}

\maketitle

\begin{quote}
\small
\textbf{\textsf{Abstract}}

\smallskip

\textbf{Purpose:} Sodium MRI is challenging because of the low tissue concentration of the  \textsuperscript{23}Na nucleus and its extremely fast biexponential transverse relaxation rate.  In this article, we present an iterative reconstruction framework using dual-echo \textsuperscript{23}Na data and exploiting anatomical prior information (AGR) from high-resolution, low-noise, \textsuperscript{1}H MR images. This framework enables the estimation and modeling of the spatially-varying signal decay due to transverse relaxation during readout (AGRdm), which leads to images of better resolution and  reduced noise resulting in improved quantification of the reconstructed \textsuperscript{23}Na images. 

\textbf{Methods:} The proposed framework was evaluated using reconstructions of 30 noise realizations of realistic simulations of dual echo twisted projection imaging (TPI) \textsuperscript{23}Na data. 
Moreover, three dual echo \textsuperscript{23}Na TPI brain data sets of healthy controls acquired on a 3T Siemens Prisma system were reconstructed using conventional reconstruction, AGR and AGRdm.

\textbf{Results:} Our simulations show that compared to conventional reconstructions, AGR and AGRdm show improved bias-noise characteristics in several regions of the brain. Moreover, AGR and AGRdm images show more anatomical detail and less noise in the reconstructions of the experimental data sets. Compared to AGR and the conventional reconstruction, AGRdm shows higher contrast in the sodium concentration ratio between gray and white matter and between  gray matter and the brain stem.

\textbf{Conclusion:} AGR and AGRdm generate \textsuperscript{23}Na images with high resolution, high levels of anatomical detail, and low levels of noise, potentially enabling high-quality \textsuperscript{23}Na MR imaging at 3T.
 
\end{quote}

\textbf{Accepted for publication in Magnetic Resonance in Medicine on 04 Nov 2023.}

\begin{multicols}{2}
  \section{Introduction}

Sodium-23 (\textsuperscript{23}Na) is the second most abundant MR-active nucleus in the human body and
plays a crucial role in ion homeostasis and cell viability.
Changes in tissue sodium concentration have been shown to provide physiologically relevant information for 
challenging brain pathologies such  such as brain tumors \cite{ouwerkerk2003,bartha2008}, 
stroke \cite{Hussain2009,Boada2012} and Alzheimer's disease \cite{Mellon2009, Mohamed2021, Haeger2022}. 
For an in-depth overview on the state-of-the-art of in vivo MR sodium imaging techniques
and applications, we refer to the recent reviews by Madelin \cite{Madelin2013}, Sha \cite{shah2016}, 
Thulborn \cite{thulborn2018} and Hagiwara \cite{hagiwara_sodium_2021}.
Despite its unique potential for providing non-invasive and physiologically relevant information, 
the development and clinical acceptance of sodium MRI has been hampered compared to proton MRI, owing to 
the unique data acquisition challenges posed by the NMR properties of the \textsuperscript{23}Na nucleus, namely: 
\begin{enumerate}
 \item lower NMR sensitivity, due to its 4x lower gyromagnetic ratio
 \item (ca. 2000x) lower tissue concentration
 \item very short (biexponential) transverse ($T_2^*$) relaxation.
\end{enumerate}
The combination of the latter two properties limits the maximum k-space 
frequency $k_\text{max}$ that can be used for acquiring images of adequate signal-to-noise ratio (SNR) 
in clinically practical imaging times (ca. 10\,min). 
Moreover, the images that are commonly obtained, suffer from spatial blurring beyond that corresponding to the nominal
spatial resolution of the acquired spatial frequency values (i.e., $1/2k_\text{max}$) due to the 
fast decay of the signal during readout. 
As discussed in \cite{stobbe2018, Haeger2022}, a direct consequence of the rather low resolution
of sodium images are severe partial volume effects (PVEs) that make accurate quantification
of local sodium concentration difficult. 
This is especially the case, when changes in local sodium concentration in the
brain are accompanied by atrophy, necessitating advanced methods for partial volume correction.

In layman's terms, one could say that compared to standard proton MR images,
due to the unfavorable NMR properties of 
\textsuperscript{23}Na, sodium MR images are ``blurry'' and ``noisy'' - similar
to images acquired in emission tomography (PET and SPECT).
To simultaneously deblur and denoise emission tomography images,
several reconstruction methods
based on high-resolution and low-noise structural (anatomical) prior images
have been proposed over the last 30 years; see, e.g.
\cite{fessler1992, lipinski1997, baete2004, bowsher2004, Vunckx2012, knoll2016, ehrhardt2016a, schramm2017}.
Similar methods have been applied in proton MR imaging in the context of
spectroscopic imaging \cite{liang1991}, extended k-space sampling \cite{haldar2008} and undersampled multicontrast MR reconstruction
\cite{Ehrhardt2016}.
In \textsuperscript{23}Na MR reconstruction, Gnahm et al. proposed to include
structural prior information from various proton MR contrasts using a anatomically-weighted 
first and second order total variation (AnaWeTV) \cite{gnahm2015}.
Based on reconstructions of simulated and real data, and in agreement with the
results obtained in articles on structure-guided reconstruction in emission tomography, 
the authors conclude that
``The approach (AnaWeTV) leads to significantly increased SNR and enhanced resolution of known
structures in the images ... Therefore, the AnaWeTV algorithm is in particular beneficial for the evaluation of tissue structures that are visible in both \textsuperscript{23}Na and 
\textsuperscript{1}H MRI.''
In 2021, Zhao et al. \cite{zhao2021} proposed a reconstruction method based on
a motion compensated generalized series model and a sparse model where anatomical
prior information from a segmented high-resolution proton image was used
for denoising and resolution enhancement.
The authors also concluded that the proposed anatomically
constrained reconstruction method substantially improved SNR and lesion
fidelity.

In this work, we propose using anatomical guidance based on the segmentation-free
directional total variation (dTV) prior not only for denoising and resolution 
enhancement of \textsuperscript{23}Na images, but also for local signal decay
estimation and compensation in a joint reconstruction framework for dual echo 
\textsuperscript{23}Na acquisitions.
In the context of multicontrast proton MRI reconstruction, Ehrhardt and Betcke \cite{Ehrhardt2016} already showed that dTV is superior compared to AnaWeTV.
Furthermore, in 2022 Ehrhardt et al. \cite{Ehrhardt2022} showed that dTV is
a promising prior for denoising and resolution enhancement of hyperpolarized carbon-13 MRI, which motivated us to also use dTV for the reconstruction
of dual echo sodium MR data.

The rest of the article is structured as follows:
In the next section, we explain the theory behind dual echo \textsuperscript{23}Na reconstruction, including local $T_2^*$ decay estimation and modeling, as well
as anatomical guidance through directional total variation before evaluating our
framework based on simulated and experimental data from healthy human volunteers.

\section{Theory}

\subsection{Conventional iterative single-echo sodium MR image reconstruction}

Assuming additive uniform Gaussian noise on the measured complex-valued (non-Cartesian) single-echo MR raw data $d \in \mathbb{C}^m$, 
we can define the conventional maximum a posteriori single echo data MR reconstruction 
problem via the optimization problem
\begin{equation} \label{eq:basic_problem}
   \argmin_u \Biggl( \frac{1}{2} \sum_{l=1}^m | F_{k(t_l)} u - d_l |^2 + \beta R(u) \Biggr)  \ ,
\end{equation}
with $u \in \mathbb{C}^n$ being the complex (sodium) MR image to be reconstructed,
$R: \mathbb{C}^n \to \mathbb{R}$ being a regularization functional weighted by the
real-valued scalar $\beta$. 
In \eqref{eq:basic_problem} $| \cdot |$ denotes the absolute value of a complex number, 
and we define 
$F_{k(t)}: \mathbb{C}^n \to \mathbb{C}$ as the Fourier encoding operator mapping a
complex-valued input image $w$ to a single point in k-space ($k_x(t)$,$k_y(t)$,$k_z(t)$) via
\begin{equation} \label{eq:fwd_model_no_decay}
   F_{k} w = \sum_{j=1}^n w_j \exp \Bigl(-i (k_x x_j + k_y y_j + k_z z_j) \Bigr) \ ,
\end{equation}
where ($x_j$,$y_j$,$z_j$) are the spatial coordinates of voxel $j$.
Given a k-space readout trajectory $(k_x(t), k_y(t), k_z(t))$ as a function
of time $t$, 
$F$ can be evaluated using the discrete-time Fourier
transform or approximated by the non-uniform fast Fourier transform (NUFFT) \cite{fessler2003}.
Note that if the readout time is not short compared to the
spatially-varying transverse relaxation times $T_2^*(x,y,z)$,
k-space data points acquired ``late'' after excitation 
suffer from apodization caused by $T_2^*$ relaxation of the signal.
For typical non-Cartesian readouts that move outwards starting at the center of k-space,
that implies damping of the high frequency data points leading to a spatially-varying loss of resolution when
optimizing \eqref{eq:basic_problem}.

\subsection{Sodium MR reconstruction including transverse relaxation modeling}

If the spatially-varying transverse relaxation is known, 
it can in principle be included in the forward model \eqref{eq:basic_problem} to
compensate for the added blurring caused by the transverse relaxation.
Assuming a generic known transverse relaxation function $d(x,y,z,t) \to [0,1]$ 
(e.g. a multi-exponential decay) and ignoring the effects of main magnetic field inhomogeneities on the k-space trajectory \cite{noll1991,noll1992} 
because of sodium's much lower gyromagnetic ratio, we can rewrite the
Fourier forward model \eqref{eq:basic_problem} into
\begin{equation} \label{eq:fwd_model_w_decay}
\begin{split}
   \tilde{F}_{k(t_l)} w = \sum_{j=1}^n & w_j \, d(x_j,y_j,z_j,t_l) \\
     & \exp \Bigl(-i (k_x(t_l) x_j + k_y(t_l) y_j + k_z(t_l) z_j) \Bigr).
\end{split}
\end{equation}
In case the transverse relaxation is modeled by an effective spatially-varying
monoexponential decay, we obtain 
\begin{equation} \label{eq:fwd_model_me}
\begin{split}
   \tilde{F}_{k(t_l)} w = \sum_{j=1}^n & w_j \, \exp\Bigl(-\frac{t_l}{T_2^*(x_j,y_j,z_j)}\Bigr)  \\
     & \exp \Bigl(-i (k_x(t_l) x_j + k_y(t_l) y_j + k_z(t_l) z_j) \Bigr) \ ,
\end{split}
\end{equation}
where $T_2^*(x_j,y_j,z_j)$ denotes the effective monoexponential transverse
relaxation time at voxel $j$.
Note, however, that in practice the transverse relaxation function $d(x,y,z,t)$ is
unknown, since it depends, among other things, on the type of tissue and 
inhomogeneities of the field $B_0$.

One way to obtain information about the transverse relaxation function $d$ is
to perform multiple acquisitions at different echo times.
A ``naive'' way to extract $d$ would be to independently reconstruct
all data sets by optimizing \eqref{eq:basic_problem} followed by a voxel-wise
fit of the decay model.
However, this approach is severely hampered by the inherent low SNR of
the sodium MR signal and the finite time available for data acquisition.

As a better alternative, we propose to jointly estimate a simplified
spatially-varying decay model during reconstruction of dual-echo
data, as explained in the next section.

\subsection{Dual-echo sodium MR image reconstruction including joint T2* estimation}

\begin{figure}
    \centering
    \includegraphics[width=1.0\columnwidth]{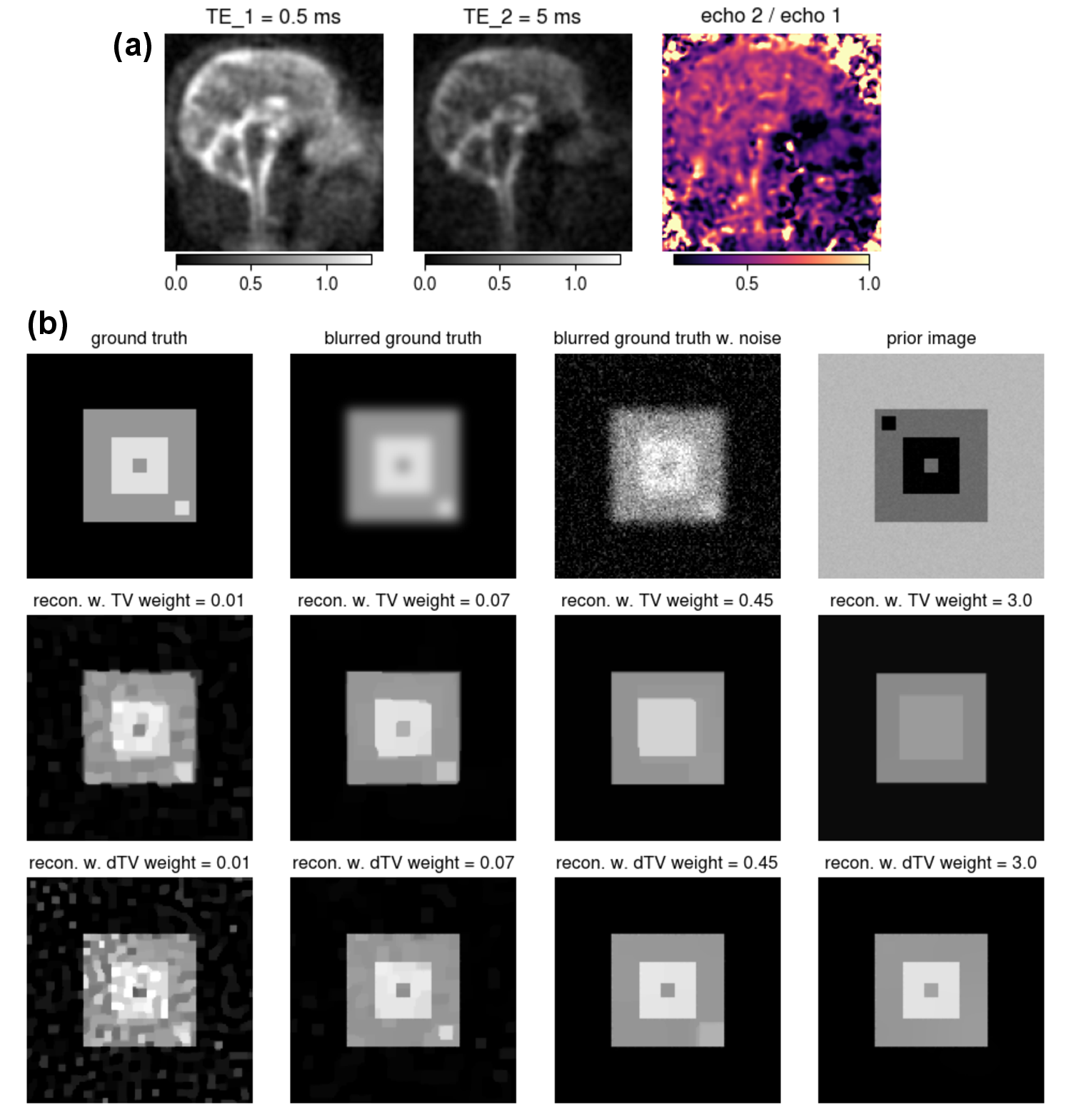}
    \caption{(a) (left): sagittal slice of \textsuperscript{23}Na twisted projection imaging (TPI) acquisition with $T_{E1}$ = 0.5\,ms
    (middle): same as left with  $T_{E2}$ = 5.0\,ms
    (right) ratio between $T_{E2}$ and $T_{E1}$ image showing spatially-varying rapid (e.g. regions close to nasopharynx, oropharynx and deep paranasal sinuses) and slower (e.g. CSF in ventricles) $T_2^*$ relaxation.
    (b) Illustration of structure-guidance with directional TV applied
    to an image deblurring and denoising problem.
    (Top row) ground truth, blurred ground truth, blurred ground truth
    with added noise (data to be reconstructed), and structural prior image.
    (Middle row) reconstructions using total variation (TV) as regularizer.
    (Bottom row)  reconstructions using directional TV (dTV) as regularizer.
    The level of regularization is increasing from left to right.
    In contrast to TV, dTV avoids smoothing across edges that are
    present in the prior image.
}
    \label{fig:dual_echo_dtv}
\end{figure}

The availability of two sodium MR acquisitions of the same subject at different echo times
(e.g. 0.5\,ms and 5\,ms) allows extracting information about the spatially-varying
transverse relaxation as illustrated in \ref{fig:dual_echo_dtv}(a).
For convenience, we re-write the monoexponential decay model used in \eqref{eq:fwd_model_me} 
into the power function
\begin{equation}
  \exp\Bigl(-\frac{t_l}{T_2^*(x_j,y_j,z_j)}\Bigr) = r(x_j,y_j,z_y)^{\sfrac{t_l}{\Delta T_E}} \ ,
\end{equation}
where $\Delta T_E$ is the difference between the first and the second echo time and
$r \in [0,1]^n$ being an exponential transformation of the spatially-varying
effective monoexponential $T_2^*$ map defined via
\begin{equation}
    r_j = \exp \Biggl( -\frac{\Delta T_E}{T_2^*(x_j,y_j,z_j)} \Biggr) \ .
    \label{eq:r_T2}
\end{equation}

Combining the data fidelity terms of both readouts and using the mappings
\begin{align}
   A_{1,l}(r) &=  F_{k(t_l)} \diag( r^{\sfrac{t_l} {\Delta T_E}} ) \label{eq:fwd_r1} \\
   A_{2,l}(r) &=  F_{k(t_l)} \diag( r^{\sfrac{t_l} {\Delta T_E} + 1} ) \label{eq:fwd_r2} \ ,
\end{align}
leads to the following dual-echo reconstruction problem with joint decay estimation
\begin{equation} \label{eq:problem}
(u^\dagger, r^\dagger) \in \argmin_{u,r} \mathcal{L}(u,r)
\end{equation}
\begin{equation} \label{eq:cost}
\begin{split}
    \mathcal{L}(u,r) = \Biggl( &\frac{1}{2} \sum_{l=1}^m | A_{1,l}(r)  u - d_l^{T_{E1}} |^2 + \\ 
   &\frac{1}{2} \sum_{l=1}^m | A_{2,l}(r)  u - d_l^{T_{E2}} |^2 + \\
   &\beta_u R_u(u)  + \beta_r R_r(r) + \chi_{(0,1]} (r) \Biggr) \ ,
\end{split}
\end{equation}
where $u \in \mathbb{C}^n$ is the complex sodium image to be reconstructed,
$r \in \mathbb{R}^n$ is the real ``ratio decay'' image to be jointly estimated,
$\chi_{(0,1]}$ is the characteristic function of the convex set $\{r \mid 0 < r_j \leq 1  \ \forall j \}$, and
$d_l^{T_{E1}}$ and $d_l^{T_{E2}}$ denote the $l$-th data point acquired at readout time $t_l$ corresponding to the 3D k-space point
$k_l = k(t_l) = (k_x(t_l), k_y(t_l), k_z(t_l))$ defined by the k-space trajectory from the
acquisitions using the first ($T_{E1}$) and second echo time ($T_{E1} + \Delta T_E$), respectively.
Note that in this work, we approximate the biexponential transverse relaxation using an effective
monoexponential because: (i) acquisitions at two echo times were available and (ii) fitting
a biexponential is less stable even if data from more echo times would be available.
We also assume that data are acquired using a single channel volume head coil
with a constant coil sensitivity.
However, an extension of the data fidelity terms to multi-channel coil data with known
sensitivity maps is straightforward.
Finally, $R_u : \mathbb{C}^n \to \mathbb{R}$ and $R_r : \mathbb{R}^n \to \mathbb{R}$ are regularization
functionals that are needed since equation \eqref{eq:problem} is ill-posed because
the acquired sodium MR raw data have low SNR, particularly at high spatial frequencies
due to apodization caused by the fast transverse relaxation.

\subsection{Structure-guided regularization via directional total variation}

When reconstructing low-resolution data and assuming that the image to be
reconstructed is structurally similar to a prior image with higher resolution, 
it is reasonable to use regularization functionals that incorporate structure-guided
information.
Over the last couple of years, many of those regularization functionals have been proposed - see, e.g. the review book chapter \cite{Ehrhardt2021}.
A simple but powerful example is ``directional total variation`` (dTV) proposed in \cite{Ehrhardt2016},
which was already applied to the reconstruction of multicontrast proton MR \cite{Ehrhardt2016} and also
hyperpolarized carbon-13 MR \cite{Ehrhardt2022}.
dTV can be seen as a structural extension of standard total variation in the presence of
a (high resolution and low noise) prior image $v$.

Total variation can be defined as 
\begin{equation}
    \text{TV}(u) = \sum_{j=1}^n | \nabla u_j  | \ ,
\end{equation}
where $\nabla$ is the three-dimensional gradient operator mapping from $\mathbb{C}^n$ to $\mathbb{C}^{3n}$.
In other words, to compute $\text{TV}(u)$, we first evaluate the gradient operator applied to the
image $u$ at every voxel $j$ (e.g., by using the finite forward differences in all three directions)
which results in a three-element gradient vector for every voxel $j$.
Subsequently, we calculate a norm of the gradient vector (e.g. the L1 or L2 norm) at every voxel and
finally sum all norms.

Based on this definition, Ehrhardt et al. \cite{Ehrhardt2016} defined directional total variation
as
\begin{equation}
    \text{dTV}(u) = \sum_{j=1}^n | P_{\xi_j} \nabla u_j  | \ ,
\end{equation}
where $P_{\xi_n}: \mathbb{C}^3 \to \mathbb{C}^3$ is a projection
operator defined via
\begin{equation} \label{eq:proj_xi}
    P_{\xi_j} g = g - \langle \xi_j, g \rangle \xi_j  \ ,
\end{equation}
and $\xi_n$ is a normalized gradient vector-field derived from a structural prior image $v$ via
\begin{equation}
  \xi_j = \frac{\nabla v_j}{ \sqrt{| \nabla v_j |^2 + \eta^2}} \ ,
\end{equation}
where $\eta$ is a scalar that determines whether a gradient is considered
large or small.
$\eta$ can be chosen based on the noise-level in the prior image $v$, such that
gradients due to noise are suppressed.
Note that in the limit case $| \nabla v_j | \to 0$ (locally flat prior image), 
dTV reduces to TV. 
In the other limit case $| \nabla v_j | \gg \eta$, 
$\xi_j = \nabla v_j / |\nabla v_j|$ and hence $|\xi_j| = 1$,
which means that $P_{\xi_j}$ applied to a gradient vector $g$ subtracts the
component of $g$ that is parallel to $\xi_j$ from $g$, or in other words,
results in the component of $g$ that is orthogonal to $\xi_j$.
Hence, instead of penalizing the norm of the ``complete'' gradient vector
as in TV,
dTV ``only'' penalizes the norm of the components of the gradient vectors that are perpendicular to the vectors of the joint gradient field.
Consequently, a parallel or anti-parallel orientation of the gradient
of the image to be reconstructed relative to the gradient field from
the prior image is encouraged, independently of the magnitude of the
joint gradient field (provided that $| \nabla v_j | \gg \eta$).
A simplified denoising and deblurring example illustrating the difference between using TV and the  structure-guided dTV prior is shown in Fig.~\ref{fig:dual_echo_dtv}(b).


\section{Methods}

To investigate our proposed structure-guided sodium MR reconstruction framework, which includes joint signal 
decay estimation and modeling, we performed reconstructions of simulated and experimental data from healthy volunteers.
In this section, we first describe the simulation setup and the acquisition and reconstruction of the experimental data, followed by a detailed explanation of our
strategy and implementation to solve the large-scale optimization (reconstruction) problem \eqref{eq:problem}.

\subsection{Simulation experiments}

The performance of the proposed sodium reconstruction framework was first investigated
using data simulated based on the brainweb phantom \cite{Collins1998}.
A high-resolution ground truth sodium image was generated using the tissue 
class label images of the brainweb phantom (subject 54). 
Fixed total sodium concentrations as well as short and long
biexponential $T_2^*$
times were assigned to each tissue class; see Tab.~\ref{table:brainweb}.
A spherical white matter lesion without corresponding edges in the structural prior
T1 image was added to the ground truth TSC image to investigate structural mismatches.
Dual-echo k-space data ($T_{E1} = 0.455\,\text{ms}$ and $T_{E2} = 5\,\text{ms}$)
including biexponential $T_2^*$ decay were simulated as a weighted sum
of the forward models in equation \eqref{eq:fwd_model_me} using the twisted projection imaging (TPI) readout
described in the next section and the simulated short (60\%) and long $T_2^*$ images (40\%).
Subsequently, uniform Gaussian noise, reflecting the SNR observed in real data, 
was added to the real and imaginary part of the simulated k-space data.

In total, 30 noise realizations were generated and reconstructed using the following
settings:
\begin{enumerate}
    \item ``Conventional'' reconstruction (\textbf{CR}) of the first echo data set by solving equation \eqref{eq:basic_problem}
          using a smooth non-structural ``quadratic difference'' prior $R(u) = \| \nabla u \|_2^2$ and a 
          forward model that neglects signal decay during readout.
          This reconstruction method is very similar to a (filtered) inverse Fourier transform 
          reconstruction of gridded k-space data, but has the advantage of correctly weighting the
          noise in the data fidelity term.
    \item ``Anatomically-guided'' reconstruction (\textbf{AGR}) of the first echo data set by solving equation \eqref{eq:basic_problem} using the structural dTV prior $R_u(u) = \text{dTV}(u)$ and a forward model that neglects signal decay during readout.
    The joint normalized gradient field $\xi$ needed for dTV 
    was derived from the high-resolution proton T1 image of the brainweb
    phantom.
    \item ``Anatomically-guided'' reconstruction of the combined first and second echo data set 
          including joint signal decay estimation and modeling (\textbf{AGRdm}) by solving equation \eqref{eq:problem}
          using the structural dTV prior\footnote{To obtain a differentiable prior functional $R_r(r)$, a modified dTV functional penalizing the squared L2 norm of the projected gradient was used.} and a monoexponential for signal decay during readout in \eqref{eq:fwd_r1}
          for $u$ and $r$.
\end{enumerate}
To study the potential bias introduced by modeling the biexponential decay using a simplified and estimated monoexponential model, simulated noise-free data sets were also reconstructed with AGRdm using the known biexponential model and compared to AGRdm using the estimated monoexponential model and AGR without decay model.
All reconstructions used a grid size of 128x128x128 and were run for various levels
of regularization ($\beta_u$ and $\beta_r$).
The quality of all reconstructions was 
assessed in terms of regional bias-noise curves evaluated in different anatomical regions of
interest (ROIs) that were defined based on grey matter, white matter and
CSF compartments of the brainweb phantom which were further
subdivided based on a freesurfer segmentation of the brainweb T1 image.
Regional bias was quantified by first calculating the regional mean value in each ROI, followed
by averaging across all noise realizations and subtracting the known ground truth values.
Regional noise was assessed by first calculating the voxel-wise standard deviation image
across noise realizations, followed by regional averaging across each ROI.
Moreover, regional reconstruction quality was assessed by calculating the root-mean-square
error in each ROI for all reconstructions across all noise realizations.
In addition, the mean effetive monoexponential $T_2^*$ times estimated with AGRdm were
quantified in cortical grey matter, white matter and in the ventricles using the estimatated
ratio images $r$ and Eq.~\eqref{eq:r_T2}.

\begin{table}
\centering
\begin{tabular}{c c c c c} 
 \hline
  & GM & WM & CSF & lesion\\ [0.5ex] 
 \hline
 TSC (arb. units) & 0.6 & 0.4 & 1.5 & 0.6 \\
 short $T_2^*$ (ms) & 3 & 3 & 50 & 3 \\
 long $T_2^*$ (ms) & 20 & 18 & 50 & 18 \\ [1ex] 
 \hline
\end{tabular}
\caption{Total sodium concentration (TSC), and short and long $T_2^*$ values for
grey matter (GM), white matter (WM), and cerebrospinal fluid (CSF) used in
brainweb-based data simulation.}
\label{table:brainweb}
\end{table}

\subsection{In vivo experiments}

In addition to the simulation experiments,
sodium MR raw data of three healthy controls (60yr male, 65 yr female, 42yr female)
acquired in a previous study approved by
the local IRB (New York University Grossman School of Medicine) were reconstructed.
In this study, two single-quantum sodium images with different echo times
were acquired on a 3T
Siemens Prisma scanner with a dual-tuned (\textsuperscript{1}H-\textsuperscript{23}Na) volume head coil
(QED, Cleveland, OH) that was also used for shimming. 
A TPI sequence \cite{Boada1997} was used
for data collection. The TPI parameters were:
rectangular RF pulse of 0.5\,ms duration, flip angle 90$^\circ$, field of view (FOV) 220\,mm, 
matrix size 64, TPI readout time 36.32\,ms, total number of TPI projections 1596, 
P = 0.4 (the fraction of $k_\text{max}$ where the transition from radial to
twisted readout occurs), TR = 100\,ms, 
TE1/TE2 = 0.5/5\,ms, 6 averages for TE1 acquisition, 4 averages for TE2 acquisition, resulting in a total acquisition time of 
15\,min\,59s / 10\,min 39\,s for the first and second echo, respectively.
In addition, a high-resolution proton MPRAGE image was acquired using
a Siemens standard 20 channel head/neck coil.
CR, AGR and AGRdm reconstructions were performed using the regularization parameters
$\beta = 0.1$ (CR), $\beta = 0.003$ (AGR) and $\beta_u = 0.003$, $\beta_r = 0.3$
(AGRdm) and used a grid size of 128x128x128 and 2000 image updates.
Since the proton MPRAGE was acquired with a different coil which required patient repositioning,
the proton MPRAGE was rigidly aligned to CR using simpleITK and mutual information as as loss function
before running AGR and AGRdm.

To quantify local sodium concentrations, Freesurfer v7.3 was used to segment and parcellate the proton MPRAGE image.
Subsequently, mean sodium concentrations were calculated in cortical gray matter,
white matter, for all reconstructions.
The mean sodium concentration in vitreous humor of AGRdm (145 mmol/L) was used as internal reference to scale all sodium reconstructions from arbitrary units to mmol/L.
In the absence of external sodium concentration calibration standards, sodium concentration ratios between cortical gray matter and white matter, and between cortical gray matter and a brainstem ROI eroded by 4\,mm were also calculated, similar to the relative regional sodium quantification proposed in \cite{Haeger2022}.

\subsection{Optimization / reconstruction algorithm}

Optimizing equation \eqref{eq:cost} is nontrivial, since the cost function $\mathcal{L}(u,r)$ is not jointly convex nor bi-convex.
Therefore, we propose to optimize \eqref{eq:cost} using the alternating update scheme
\begin{align}
 u^{(n+1)} &= \argmin_u \mathcal{L}(u, r^{(n)}) \label{eq:update_u} \\   
 r^{(n+1)} &= \argmin_r \mathcal{L}(u^{(n+1)}, r)  \label{eq:update_r} \ ,
\end{align}
after careful initialization of $u$ and $r$. 
Note that the forward mappings $A_{1/2,i}$ are linear in
$u$ and that subproblem \eqref{eq:update_u} is convex in $u$, 
if the regularization functional $R_u$ is convex, which is the case for TV or dTV.
This, in turn, allows applying any standard convex optimizer to solve
\eqref{eq:update_u} - and also \eqref{eq:basic_problem}.
In this work, we use the well-studied primal-dual hybrid-gradient (PDHG)
algorithm by Chambolle and Pock \cite{Chambolle2011} to solve \eqref{eq:update_u},
which is efficient since the proximal operators of the data fidelity and the 
TV/dTV term are ``simple''.

Solving subproblem \eqref{eq:update_r} is more complicated, since our forward
model is non-linear in $r$ and, moreover, \eqref{eq:update_r} is not convex in $r$.
To minimize \eqref{eq:update_r}, we apply the projected gradient descent update
\begin{equation}
   r^{(n+1)} = \mathcal{P}_{\chi_{(0,1]}} \Bigl( r - \alpha \, \nabla_r \mathcal{L}(u^{(n+1)}, r) \Bigr)  \ ,
\end{equation}
with $\mathcal{P}_{\chi_{(0,1]}}$ being the projection operator onto $\chi_{(0,1]}$,
the step size $\alpha = 0.3$, and the gradient 
\begin{equation}
  \begin{split}
   \nabla_r &\mathcal{L}(u, r) = \\
   &\sum_{l=1}^m \tau_l \, \diag(r)^{\tau_l-1} \operatorname{Re}\bigl( \diag({u^*}) A_{1,l}^* (A_{1,l}u - d_l^{T_{E1}}) \bigr) + \\ 
   &\sum_{l=1}^m (\tau_l+1) \, \diag(r)^{\tau_l} \operatorname{Re}\bigl( \diag(u^*) A_{2,l}^* (A_{2,l}u - d_l^{T_{E2}}) \bigr) + \\ 
   &\beta_r \nabla_r R_r(r) \ ,
  \end{split}
\end{equation}
where $\tau_l = t_l / \Delta T_E$ and $A^*$ being the Hermitian adjoint of
operator $A$.
 
For the initialization of $u$ and $r$, we first perform independent reconstructions
of the data from the two acquisitions without
modeling signal decay during readout (independent solutions of
\eqref{eq:basic_problem} for $d^{T_{E1}}$ and $d^{T_{E2}}$ using 2000 PDHG iterations).
The ``decay ratio image`` $r$ is subsequently initialized with the ratio of
the magnitude images of the two reconstructions, and $u$ is initialized with the
independent reconstruction of $d^{T_{E1}}$.
The complete reconstruction is summarized in Algorithm~\ref{alg:joint_est}.
For all reconstructions, 20 ``outer'' and 100 ``inner'' iterations
(i.e. 100 PDHG iterations in step 8 and 100 projected gradient descent
iterations in step 10) were used
in Algorithm~\ref{alg:joint_est} resulting in 2000 updates of $u$ and $r$.
All Fourier, gradient and proximal operators needed in Algorithm~\ref{alg:joint_est}
were implemented in sigpy v0.1.25 which also provides an
implementation of PDHG.

\begin{algorithm*}[t]
\begin{algorithmic}[1]
\State \textbf{align anatomical prior image} $v$ to reconstruction without structural prior
\State \textbf{independent recon of 1st echo data} $u^{T_{E1}} = \argmin_u \frac{1}{2} \sum_{i=1}^m | F_{k(t_i)} u - d_i^{T_{E1}} |^2 + \beta_u R(u)$ \Comment{using PDHG}
\State \textbf{independent recon of 2nd echo data} $u^{T_{E2}} = \argmin_u \frac{1}{2} \sum_{i=1}^m | F_{k(t_i)} u - d_i^{T_{E2}} |^2 + \beta_u R(u)$ \Comment{using PDHG}
\State \textbf{initialize} $u = u^{T_{E1}}$ 
\State \textbf{initialize} $r = \text{abs}(u^{T_{E2}}) / \text{abs}(u^{T_{E1}})$ 
\State \textbf{setup joint gradient field} $\xi$ and $\mathcal{P}_\xi$ based on $v$
\Repeat
	\State $u^{(n+1)} = \argmin_u \mathcal{L}(u, r^{(n)})$ \Comment{update sodium image using PDHG}
    \Repeat
      \State $r^{(n+1)} = \mathcal{P}_{\chi_{(0,1]}} \Bigl( r^{(n)} - \alpha \, \nabla_r \mathcal{L}(u^{(n+1)}, r^{(n)}) \Bigr)$ \Comment{update decay model using projected gradient descent}
    \Until{stopping criterion fulfilled}
\Until{stopping criterion fulfilled}
\State \Return{$u$}
\end{algorithmic}
\caption{Dual echo sodium MR reconstruction with joint $T_2^*$ estimation using structure-guided regularization}
\label{alg:joint_est}
\end{algorithm*}
%

\begin{figure*}
     \centering
     \includegraphics[width=\textwidth]{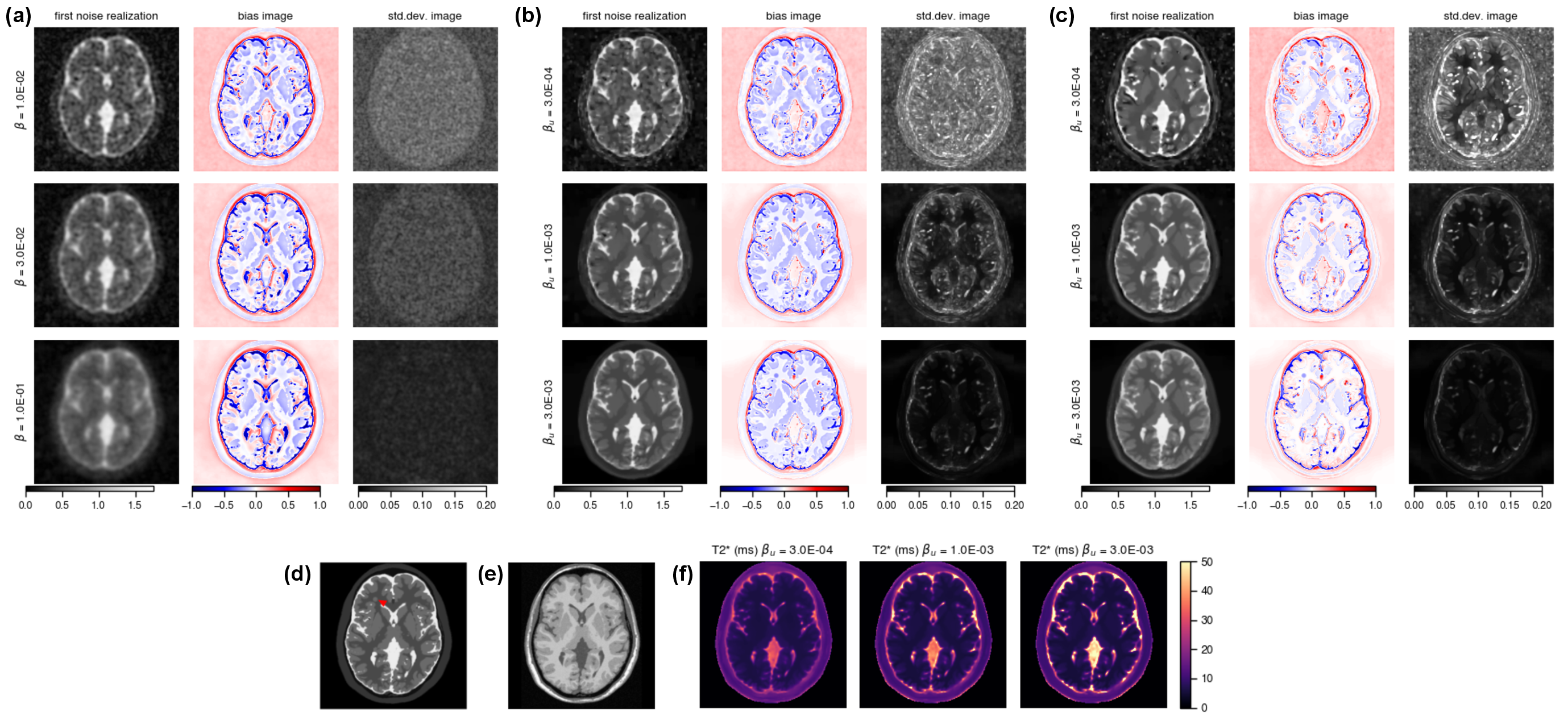}
     \caption{(a-c) Transaxial slice of first noise realization, 
              bias image (mean minus ground truth) and standard
              deviation image of the 30 simulated noise realizations for CR (a),
              AGR (b) and AGRdm (c).
              The level of regularization ($\beta$ / $\beta_u$) is increasing by a factor of 3 between rows from top to bottom to show the behaviour of the different reconstructions at different levels of regularization.
              (d) ground truth sodium image based on brainweb phantom. (e) proton T1 structural prior image used for AGR. The red arrow indicates the added lesion in the sodium ground truth images that is not present in the structural prior image.
              (f) estimated effective monoexponential $T_2^*$ time of AGRdm for different $\beta_u$
              }
     \label{fig:sim_results}
\end{figure*}

\begin{table}
    \centering
    \begin{tabular}{cc|r}
    \hline
            &           & mean estimated \\ 
    region  & $\beta_u$ & $T_2^*$ (ms)  \\ \hline
    cortial & 3e-4 & 8.6 \\
    grey    & 1e-3 & 8.4 \\
    matter  & 3e-3 & 8.3 \\ \hline
    white   & 3e-4 & 6.5 \\
    matter  & 1e-3 & 5.8 \\
            & 3e-3 & 5.7 \\ \hline
    ventricles & 3e-4 & 24.2 \\
               & 1e-3 & 30.8 \\
               & 3e-3 & 36.8 \\ \hline
    \end{tabular}
    \caption{Mean effective monoexponential $T_2^*$ times estimated in AGRdm of the simulated brainweb data in different regions of interest for different levels of $\beta_u$.}
    \label{tab:est_T2star}
\end{table}
    
\begin{figure*}
    \centering
    \includegraphics[width=1.0\textwidth]{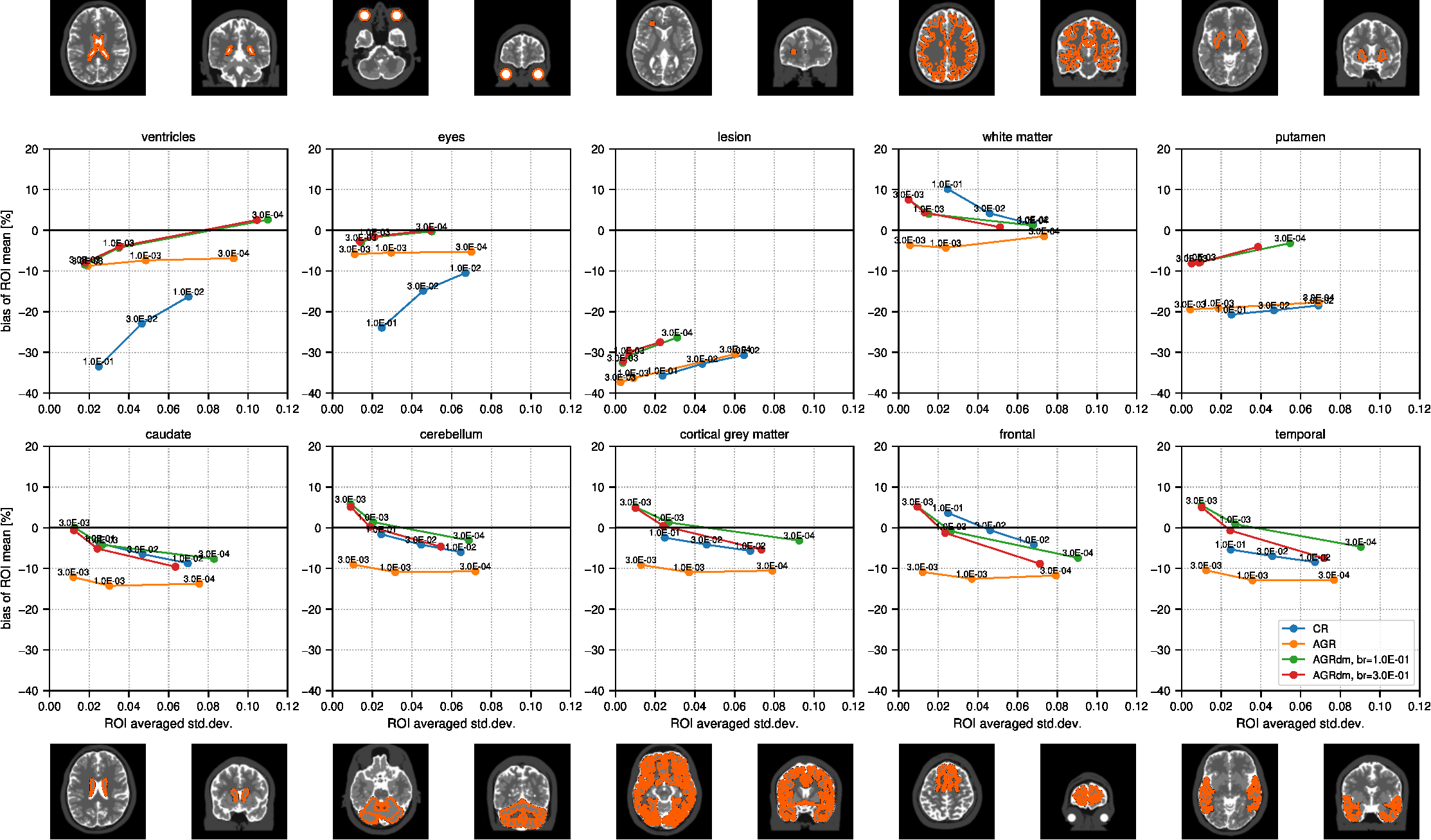}
    \caption{Regional bias noise curves calculated from 30 noise realizations from data simulated
    based on the brainweb phantom as a function of the regularization weight $\beta_u$ shown
    next to the curves. 
    Every subplot shows the result for a different region of
    interest (ROI) which are depicted above / below the subplots in a transaxial and coronal slice.}
    \label{fig:bias_noise}
\end{figure*}

\begin{figure*}
    \centering
    \includegraphics[width=1.0\textwidth]{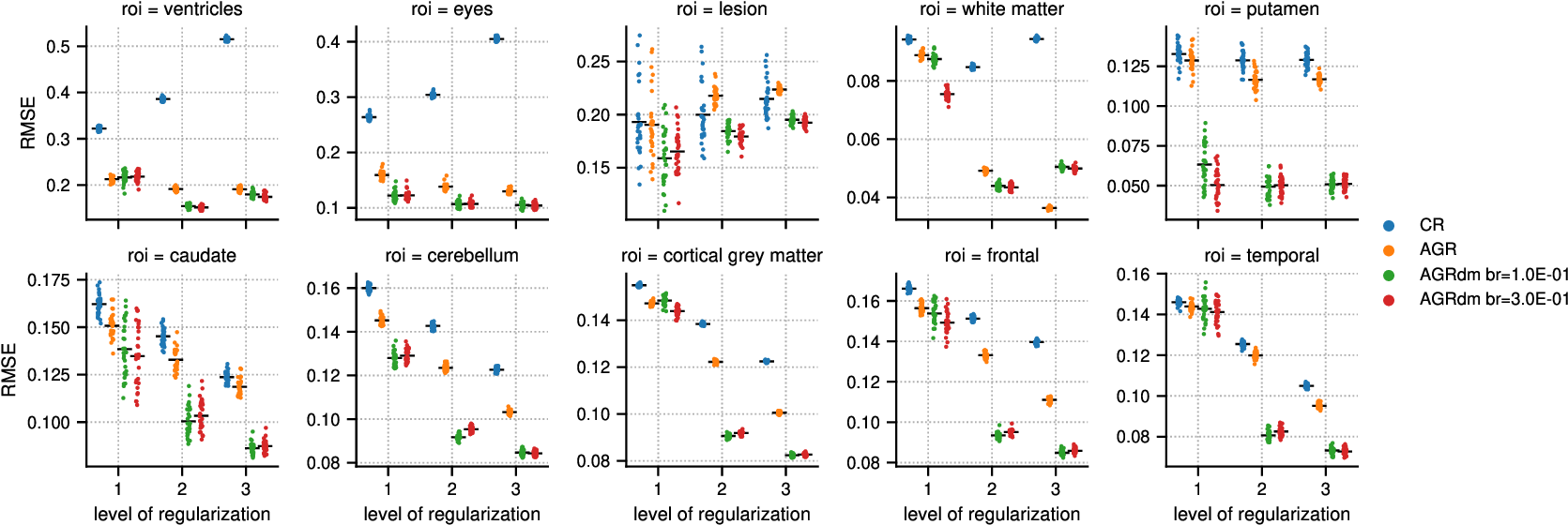}
    \caption{Regional root-mean-square error (RMSE) for all reconstructions and all 30 noise 
             noise realizations (colored dots) for three levels of regularization 
             (1 = low, 2 = medium, 3 = high). The mean regional RMSE across all noise realizations
             is indicated by black horizontal lines.}
    \label{fig:rmse}
\end{figure*}

\section{Results}
\subsection{Simulation experiments}

Figure~\ref{fig:sim_results} shows the first noise realization, the bias
(mean over all noise realizations minus ground truth) image and
the standard deviation image for the conventional reconstruction (CR) using a standard quadratic
difference prior,
the anatomically-guided reconstruction of the first echo without signal decay modeling (AGR) and
the anatomically-guided reconstruction of the data of both echos including signal decay estimation 
and modeling (AGRdm) for three different levels of image regularization ($\beta_u$).
The corresponding quantitative regional bias-noise curves as well as illustrations of all ROIs 
are shown in Fig.~\ref{fig:bias_noise}.

Comparing the image quality of the first noise realization in Fig.~\ref{fig:sim_results},
it can be seen that at the medium and high level of regularization, the boundaries between GM, 
WM and CSF are clearly visible in AGR and AGRdm, illustrating the difference in GM and WM 
sodium concentration which is hardly visible in CR.
Moreover, both AGRs show much better separation of CSF contained in
sulci between the GM gyri.

In the conventional reconstruction, higher levels of regularization cause loss of
resolution, accompanied by more severe partial volume effects (PVE) leading to
more negative bias in the ventricles, the eyes and the putamen, which are all
structures surrounded by tissues with lower sodium concentration as shown in the top
row of Fig.~\ref{fig:bias_noise}.
In white matter, which is surrounded by regions with higher sodium concentration, the same effect leads to positive bias.
In all those ROIs, AGRdm shows the least amount of bias at a given noise level.

In the cortical gray matter ROIs, shown in the bottom row of  Fig.~\ref{fig:bias_noise},
the situation is more complex.
In those regions, AGRdm shows has the best bias vs noise trade-off, whilst AGR suffers from
consistent negative bias that is bigger than the bias seen in CR.
This is because, in cortical gray matter, the positive bias due to spill-over from CSF, the 
negative bias due to spill-in to white matter and the signal loss due to
$T_2^*$ decay seem to cancel for CR.
In AGR without decay modeling, where the spill-over from CSF is strongly reduced,
the negative bias of around -10\% due to $T_2^*$ decay becomes noticeable and comparable
to the value one would expect given an echo time of 0.455\,ms, and short/long $T_2^*$ times
of 3/20\,ms.

From the standard deviation images in Fig.~\ref{fig:sim_results} it is obvious that the
variability in the conventional reconstructions is very homogeneous across the image, whereas the variability in the structure-guided reconstruction is more
concentrated around tissue boundaries.
The bias images in Fig.~\ref{fig:sim_results} and the
bias noise plot of the lesion ROI show that in the added sodium white matter
lesion, which is not present in the anatomical prior image, the AGRs show
negative bias.
However, the bias noise trade-off is not inferior compared to the
conventional reconstruction.

Figure~\ref{fig:rmse} demonstrates that AGRdm shows the lowest RMSE values across all ROIs
except for white matter where AGR has slightly lower RMSE. 
In all ROIs with correct anatomical prior information, AGR and AGRdm show the lowest RMSE
values at the highest level of regularization.

Table~\ref{tab:est_T2star} shows the effective monoexponential $T_2^*$ times estimated with AGRdm in different regions of interest.
In cortical grey matter and white matter,
the estimated effective monoexponential $T_2^*$ is in between the assigned short and long
$T_2^*$ values used for the biexponential decay simulation (3/20\,ms for grey matter, 3/18\,ms for white matter).
In the ventricles, the estimated $T_2^*$ times show negative bias with respect to the true
monoexponential $T_2^*$ time for CSF used in the decay simulation (50\,ms).

Supporting Table S1 shows the regional bias of AGRdm using the known biexponential decay model compared to AGRdm using the estimated and simplified monoexponential model. Depending on the level of regularization, the bias in cortical grey matter is reduced from -5.4\% - 4.9\% for AGRdm using the estimated monoexponential model to (0.8\% - 2.3\%). A similar trend - a bias reduction by a few percentage points - is also seen in the other regions. Moreover, it can be seen that AGR without any decay model suffers from more bias in all regions except white matter at high levels of regularization.

\subsection{In vivo experiments}

\begin{figure*}
    \centering
    \includegraphics[width=1.0\textwidth]{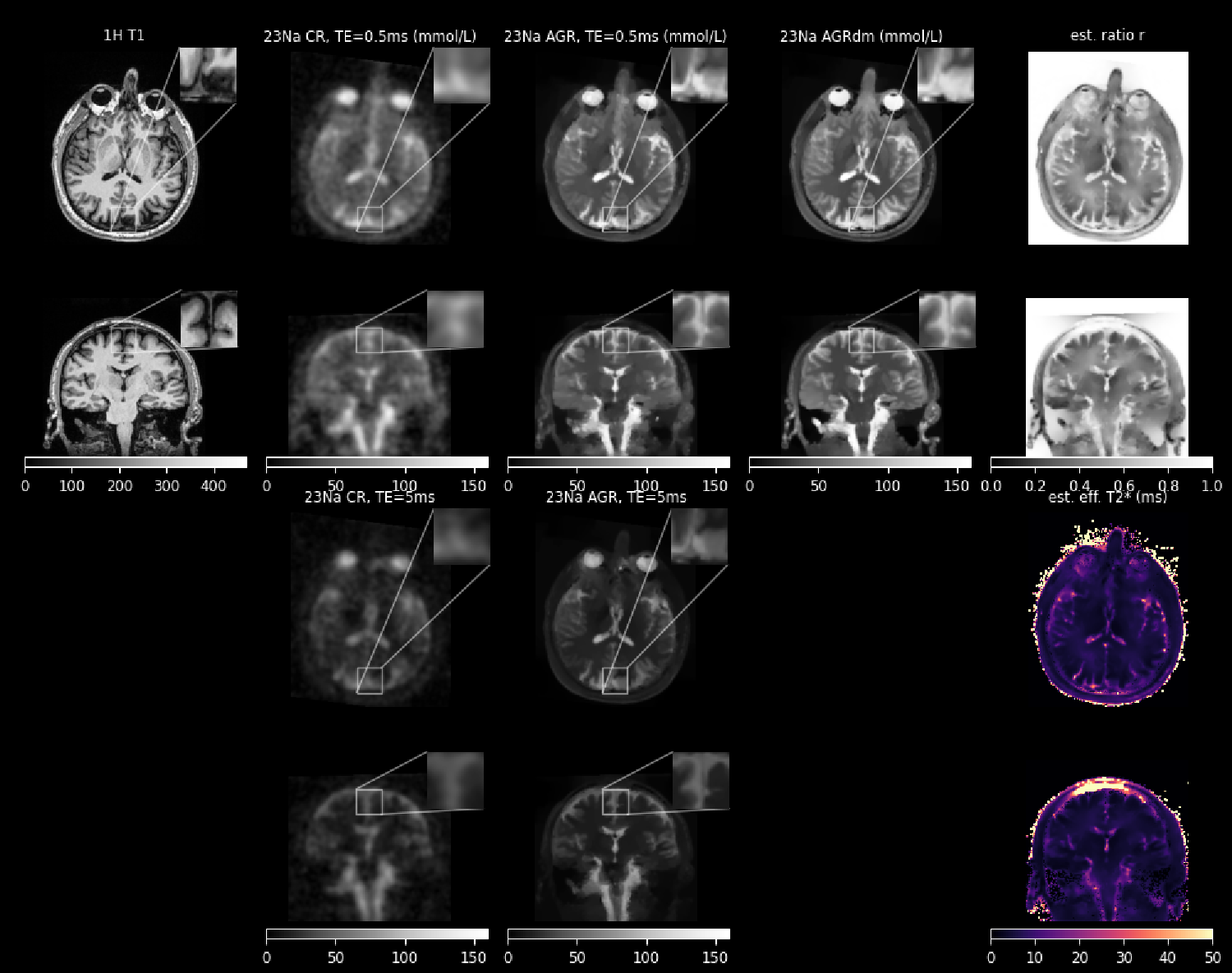}
    \caption{Dual echo sodium reconstructions of a healthy control (60yr, M) acquired on
    a 3T Siemens Prisma. (top left): \textsuperscript{1}H T1 used as anatomical prior. 
    (top 2nd from left): conventional reconstruction (CR) of the first echo 
    (top middle): anatomically-guided
    reconstruction of the first echo without signal decay modeling (AGR) (top 2nd from right):
    anatomically-guided reconstruction of both echos including signal decay
    estimation and modeling (AGRdm). 
    (top right): estimated ratio between the first and second echo ($r$) used for the monoexponential signal
    decay modeling.
    (bottom 2nd from left): conventional reconstruction (CR) of the second echo 
    (bottom middle): anatomically-guided
    reconstruction of the second echo without signal decay modeling (AGR)
    (bottom right): effective estimated monoexponential $T_2^*$ time calculated from $r$. 
    The signal intensity of all sodium images (including the 2nd echo images of CR and AGR) 
    is normalized to the mean signal intensity of AGRdm in vitreous humor (145 mmol/L) and
    shown using the same color scale.
    Due to the impact of the $T_2^*$ decay, the units of the second echo images are not
    labeled as mmol/L.    
}
    \label{fig:agr_siemens}
\end{figure*}

\begin{figure*}
    \centering
    \includegraphics[width=1.0\textwidth]{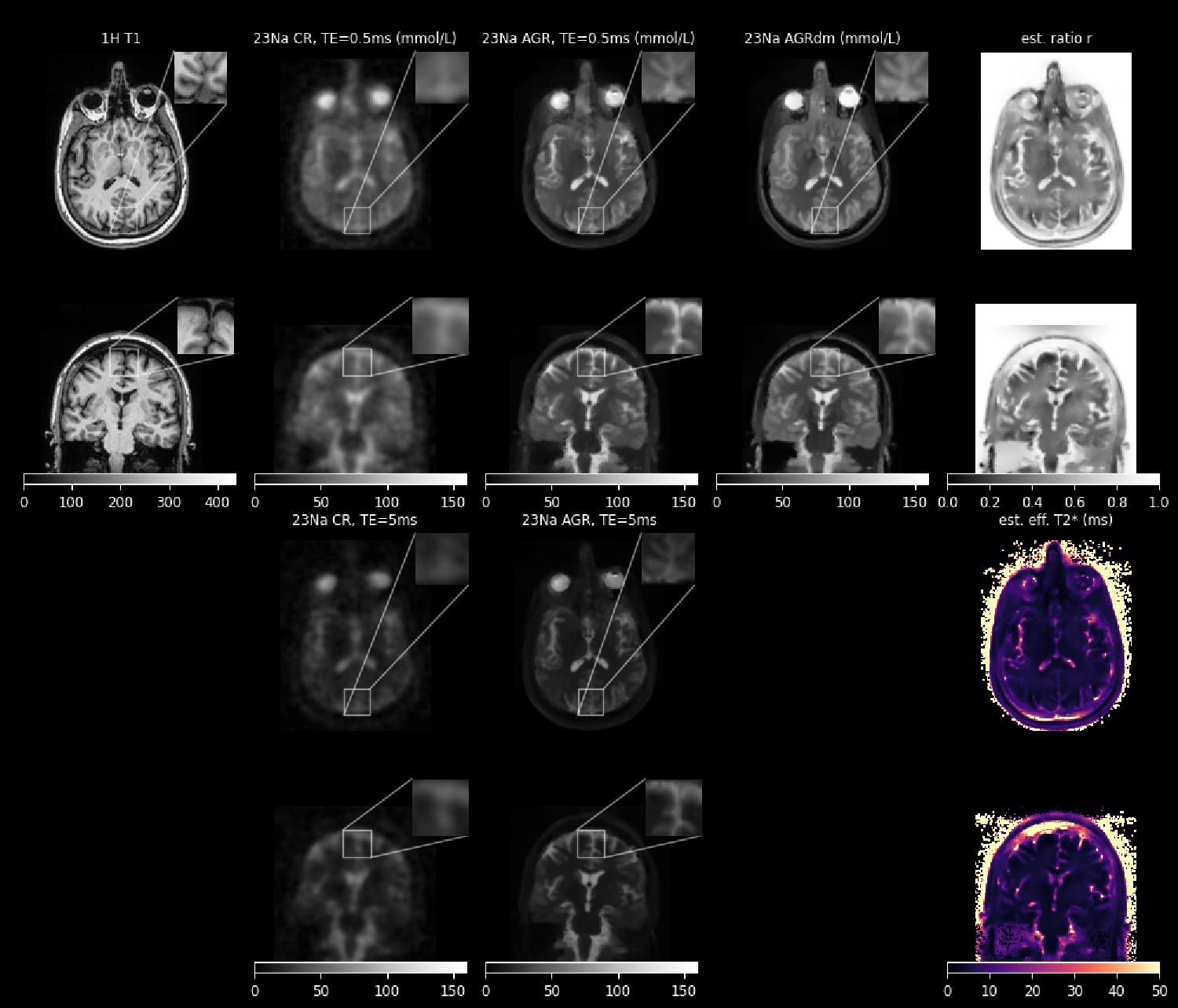}
    \caption{Same as Fig.~\ref{fig:agr_siemens} for the second healthy control (65yr, F).}
    \label{fig:agr_siemens_2}
\end{figure*}

\begin{figure*}
    \centering
    \includegraphics[width=1.0\textwidth]{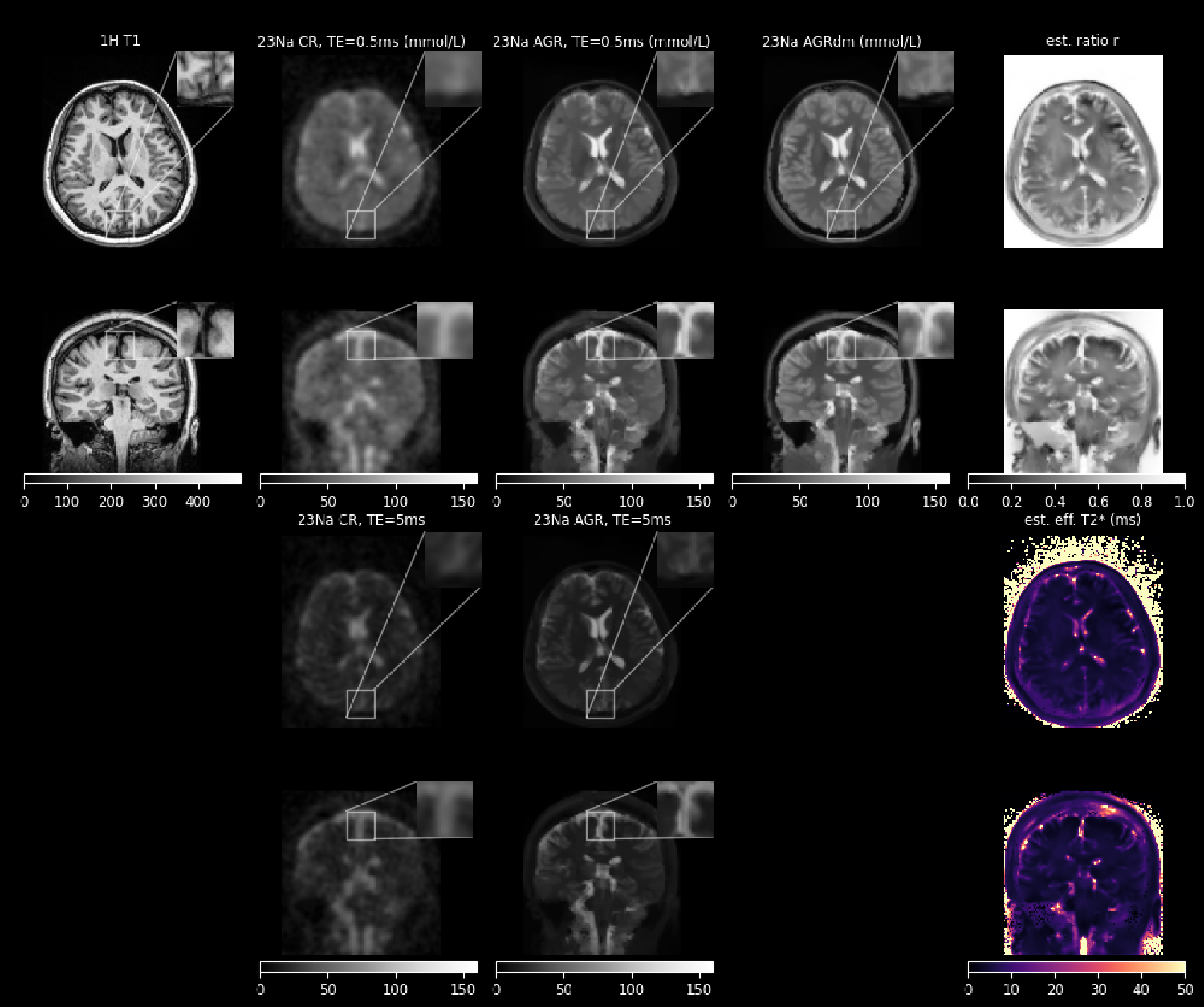}
    \caption{Same as Fig.~\ref{fig:agr_siemens} for the second healthy control (42yr, F).}
    \label{fig:agr_siemens_3}
\end{figure*}

\begin{figure*}
    \centering
    \includegraphics[width=1.0\textwidth]{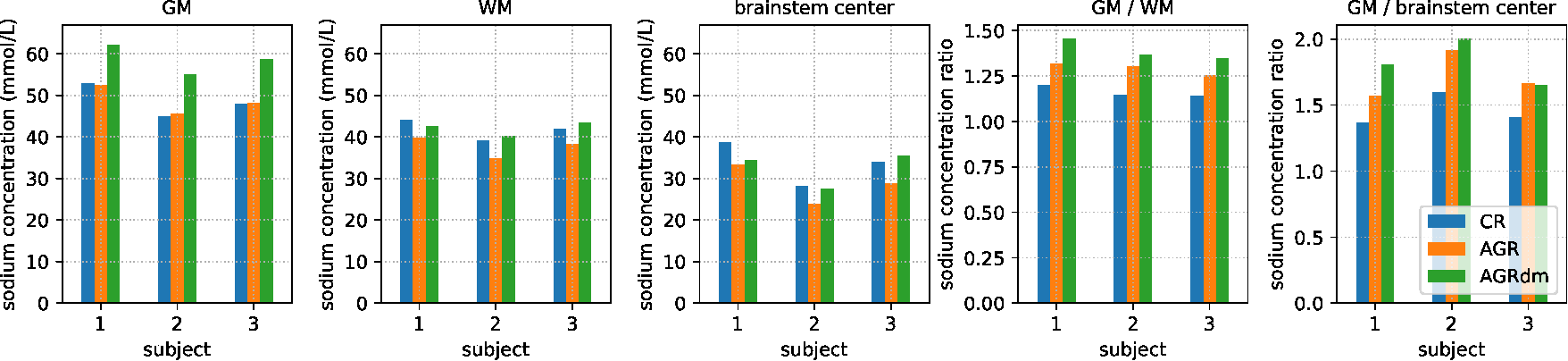}
    \caption{Regional quantification of TSC
    in GM and WM (left and middle) and the TSC GM/WM
    ratio (right) in three healthy controls for all different
    sodium reconstruction algorithms.}
    \label{fig:siemens_quant}
\end{figure*}

Figures~\ref{fig:agr_siemens}, \ref{fig:agr_siemens_2}, and \ref{fig:agr_siemens_3} show the conventional (CR) and anatomically-guided sodium reconstruction without and with signal decay estimation and modeling (AGR and AGRdm)
for the three dual echo sodium acquisitions.
In all three cases, the boundaries between GM, WM and CSF are much better
defined in the anatomically-guided reconstructions, leading, e.g., to a clearer separation between the sodium concentration in
GM and WM and also between the CSF in the sulci and cortical gray matter.
Moreover, within WM and GM, both AGRs are less noisy compared to the conventional reconstruction.
The estimated decay ratio image (the exponential transformation of the effective
estimated monoexponential $T_2^*$ time) clearly shows the relatively slow
signal decay in CSF ($r$ close to 1) and faster decay in GM and WM.
Figure~\ref{fig:siemens_quant} shows a comparison of the sodium concentration of GM and WM in the cortical region obtained with the three reconstructions in all three cases.
The higher cortical GM to WM contrast of AGRdm that can be also clearly 
seen in Fig.~\ref{fig:agr_siemens} is also confirmed in the plot of cortical GM to WM sodium concentration displayed in the right of Fig.~\ref{fig:siemens_quant}.
Morever, AGR and AGRdm also lead to higher GM to brainstem sodium concentration ratios
which are more in line with the ratios of healthy controls reported in \cite{Haeger2022}
where partial volume correction was applied post reconstruction in image space.

\section{Discussion}

In this work, we have presented a framework for joint reconstruction
and transverse signal decay estimation and modeling of dual echo \textsuperscript{23}Na 
using anatomical guidance. 
Compared to the existing works of Zhao et al. \cite{zhao2021} and Gnahm et al. 
\cite{gnahm2015}, our method has the advantages that (i) no segmentation of
the prior image is needed, (ii) a better segmentation-free anatomical prior (dTV) is used and
(iii) that the local $T_2^*$ signal decay can be estimated and modeled during reconstruction.
The latter is important since rapid spatially-varying transverse signal decay, e.g. at tissue-air
interfaces, leads to local widening of the point-spread function (PSF), which
complicates image deblurring and also leads to spatially-varying signal loss in the first echo image acquired at a finite time after excitation (e.g. 0.455\,ms).
As already mentioned in \cite{gnahm2015,zhao2021} anatomically-guided
\textsuperscript{23}Na reconstruction could have a major impact on clinical sodium MR imaging
at 3T, since it allows to reconstruct high SNR images with a high level
of detail even at 3T enabling improved \textsuperscript{23}Na imaging
at standard field strengths.
Note that although this work focuses on sodium data acquired on 3T systems, 
application of our framework to data acquired at higher field strengths is straightforward.

A current drawback of our framework is the relatively long reconstruction time.
Using a state-of-the-art GPU, a joint reconstruction using 2000 overall updates
in Alg.~\ref{alg:joint_est} takes roughly 1h.
However, it is to be expected that these reconstruction times can be
significantly shortened by using stochastic optimization algorithms
to solve \eqref{eq:update_u}, e.g., the stochastic version
PDHG \cite{Chambolle2018} and stochastic gradient descent and by using
more efficient implementations of the NUFFT.

Another limitation of our framework is the fact that most anatomically-guided
priors (including dTV) are based on the assumption that
edges in the prior image and in the image to be reconstructed are
present at the same locations.
As shown in several works, exploiting this prior knowledge during
reconstruction can be very beneficial for joint denoising and debluring.
However, all images reconstructed with anatomical prior knowledge must
be interpreted with caution if the assumption behind the anatomical
prior is violated.
This potentially means that small structures that are not present in the
structural prior image could be lost (smoothed away) when using high levels of
regularization and, therefore, the reconstruction framework must be carefully tuned to avoid such effects.
Note, however, that Fig.~4 demonstrated that the bias versus noise trade-off of AGR and AGRdm in the sodium lesion absent in the structural prior image was not inferior to CR. This can be understood by taking into account that dTV reduces to "normal" TV if no structural prior information is available (locally flat prior image).

In this work, the structural prior proton MPRAGE was acquired
with a different coil yielding superior proton image quality compared to the dual-tuned coil. A requirement of our current implementation is that the proton MPRAGE had to be rigidly registered to the sodium CR because of patient repositioning, before running AGR and AGRdm.
According to our experience the rigid registration of the proton MPRAGE to the sodium CR is stable and accurate. 
However, small residual mis-alignments cannot be fully excluded, and, threfore,  registration of the proton MPRAGE and the sodium CR should be always verified before running AGR or AGRdm since structural priors
require accurate alignment of the structural prior image.

The bias vs. noise analysis of the simulated data has shown that AGRdm 
shows less bias compared AGR and CR.
This reduction of partial volume effects, especially in the ventricles, white matter,
and gray matter, should help to improve quantification of tissue
sodium concentration in clinical acquisitions.
The comparison of AGRdm with estimated monoexponential 
and AGRdm with known biexponential decay model, which is in not availalbe in practice, showed that using the estimated
and simplified monoexponential model only leads to a slight increase of a few percentage points in the regional bias.
Taking into account that not modeling the decay in AGR leads to much stronger bias, especially in grey matter, we are confident that estimating and using a simplified monoexponential model
during AGR leads to an overall reduction of the regional bias in tissue sodium quantification.

Since no ground truth for the local sodium concentration in the in vivo
experiments was available, we decided to quantify the gray matter to 
brainstem ratio as proposed by Haeger et al.\cite{Haeger2022} in the
analysis of 3T sodium images corrected for partial volume effects
after reconstruction of 52 patients with Alzheimer's disease and
34 controls.
Compared to the ratios calculated in the conventional reconstruction
(CR), the ratios obtained from AGR and AGRdm in the three control cases
of our work were closer to the range reported for controls by
Haeger et al. (1.6 - 2.0). 
Partial volume correction post reconstruction is another option
to reduce PVEs.
Note, however, that the PV-corrected sodium image shown in Fig.~1
of \cite{Haeger2022} shows much less anatomical detail compared to the
AGR or AGRdm images shown in this work.
A more detailed investigation of the impact of anatomically guided
reconstruction on regional sodium concentration quantification in different ROIs
in a larger patient cohort is beyond the scope
of the work and left for future research.
Note that the three human dual echo sodium data sets used in this work
were acquired for a different research project and used retrospectively.
For prospective acquisitions, it is recommended to use more averages for the acquisition of the second echo since it suffers from lower SNR due to signal decay.

Finally, further gains in quantification accuracy can be gained through the use of the proposed approach by incorporating the use of $B_1$ field
mapping (to correct for RF coil "shading") and $B_0$ inhomogeneity k-space distortion correction. As shown in \cite{Lommen2016}, the former can be efficiently carried out through time-efficient
$B_1$ mapping via the 
phase-sensitive \cite{Morrell2008} Bloch-Siegert shift \cite{sacolick2010} 
or the double-angle method \cite{insko1993}. 
The later correction, on the other hand, could be performed very efficiently using a linearized version of the $B_0$ inhomogeneity map \cite{irarrazabel1996} as these effects are typically much smaller than in conventional proton MRI due to sodium's lower gyromagnetic ratio.


\section{Conclusion}

Our proposed framework for resolution enhancement, noise suppression, and joint T2* decay estimation using dual-echo sodium-23 MR data and anatomically-guided reconstruction is capable of producing 
high SNR sodium images with high levels of anatomical detail at 3T, reducing the
negative impact of partial volume effects on regional quantification in \textsuperscript{23}Na MR images.

  \subsection*{Financial disclosure}

None reported.

\subsection*{Conflict of interest}

The authors declare no potential conflict of interests.

  \section*{Supporting information}
The following supporting information is available as part of the online article:
support\_material.pdf containing supporting Table S1.

\section*{Data Availability Statement}
The input data that support the findings of the simulation study are openly available 
in the BrainWeb database at \url{https://brainweb.bic.mni.mcgill.ca/brainweb/anatomic_normal_20.html}.
The input data of the in vivo experiments are not shared.
The source code used to generate the results of this manuscript is
available at \url{https://github.com/gschramm/sodium_mr_agr_paper}.

  \printbibliography
\end{multicols}

\end{document}


\begin{table}
    \centering
    \begin{tabular}{cc|r|r|r}
               &           & AGR & AGRdm & AGRdm \\ \hline
               &  decay    & none & estimated & true \\ 
               &  model    &      & monoexp. & biexp. \\ \hline
       region  & $\beta_u$ &  \multicolumn{3}{c}{regional bias [\%]}        \\ \hline
       ventricles  & 3e-4 & -6.9 & 2.5   & -4.3  \\           
                   & 1e-3 & -7.4 & -4.0  & -2.8  \\           
                   & 3e-3 & -8.7 & -8.0  & -0.9  \\ \hline           
       eyes        & 3e-4 & -5.3 & 0.1   & -2.5  \\                
                   & 1e-3 & -5.5 & -1.5  & -2.1  \\                
                   & 3e-3 & -5.9 & -2.7  & -1.5  \\ \hline
       lesion      & 3e-4 & -30.4 & -27.5 & -30.5  \\              
                   & 1e-3 & -36.3 & -29.9 & -31.2  \\              
                   & 3e-3 & -37.3 & -32.1 & -32.0  \\ \hline
       white       & 3e-4 & -1.4 & 0.8   & 3.0  \\        
       matter      & 1e-3 & -4.3 & 4.4   & 1.8  \\        
                   & 3e-3 & -3.6 & 7.6   & 1.4  \\ \hline
       putamen     & 3e-4 & -17.7 & -4.1  & -3.4  \\            
                   & 1e-3 & -19.9 & -8.0  & -5.9  \\            
                   & 3e-3 & -19.4 & -8.2  & -8.8  \\ \hline
       caudate     & 3e-4 & -13.7 & -9.6  & -1.4  \\             
                   & 1e-3 & -14.3 & -5.1  & -1.3  \\             
                   & 3e-3 & -12.1 & -0.6  & -0.5  \\ \hline
       cerebellum  & 3e-4 & -10.7 & -4.7  & 2.4  \\          
                   & 1e-3 & -10.9 & 0.2   & 2.8  \\          
                   & 3e-3 & -9.0 & 5.1   & 3.5  \\ \hline
       cortical    & 3e-4 & -10.5 & -5.4  & 0.8  \\
       grey        & 1e-3 & -10.9 & 0.4   & 1.5  \\
       matter      & 3e-3 & -9.1 & 4.9   & 2.3  \\ \hline
       frontal     & 3e-4 & -11.7 & -8.8  & -1.6  \\             
                   & 1e-3 & -12.5 & -1.3  & -2.0  \\             
                   & 3e-3 & -10.8 & 5.2   & -2.0  \\ \hline
       temporal    & 3e-4 & -12.8 & -7.5  & -0.8  \\            
                   & 1e-3 & -12.9 & -0.7  & -0.1  \\            
                   & 3e-3 & -10.5 & 5.0   & 1.2  \\            
    \end{tabular}
    \caption{Regional bias for AGR without decay model, AGRdm with estimated monoexponential, and AGRdm with known biexponential decay model for different levels of regularization $\beta_u$ using the simulated brainweb data.}
\end{table}